\documentclass[preprint,12pt]{elsarticle}
\usepackage{amssymb}
\usepackage{amsmath}
\usepackage{hyperref}
\usepackage{lineno}
\journal{Nuclear Instruments \& Methods in Physics Research, Section A}

\begin{document}

\begin{frontmatter}

%% Title, authors and addresses

%% use the tnoteref command within \title for footnotes;
%% use the tnotetext command for theassociated footnote;
%% use the fnref command within \author or \affiliation for footnotes;
%% use the fntext command for theassociated footnote;
%% use the corref command within \author for corresponding author footnotes;
%% use the cortext command for theassociated footnote;
%% use the ead command for the email address,
%% and the form \ead[url] for the home page:
%% \title{Title\tnoteref{label1}}
%% \tnotetext[label1]{}
%% \author{Name\corref{cor1}\fnref{label2}}
%% \ead{email address}
%% \ead[url]{home page}
%% \fntext[label2]{}
%% \cortext[cor1]{}
%% \affiliation{organization={},
%%             addressline={},
%%             city={},
%%             postcode={},
%%             state={},
%%             country={}}
%% \fntext[label3]{}

\title{Yield, noise and timing studies of ALICE ITS3 stitched sensor test structures: the MOST}

%% use optional labels to link authors explicitly to addresses:
%% \author[label1,label2]{}
%% \affiliation[label1]{organization={},
%%             addressline={},
%%             city={},
%%             postcode={},
%%             state={},
%%             country={}}
%%
%% \affiliation[label2]{organization={},
%%             addressline={},
%%             city={},
%%             postcode={},
%%             state={},
%%             country={}}

\author[a,b]{J. Sonneveld}
\ead{jory.sonneveld@nikhef.nl}
\author[a,c]{R. Barthel}
\author[d]{S. Bugiel}
\author[d]{L. Cecconi\fnref{fn1}}
%\author[b]{D. Dannheim}
\author[d]{J. De Melo\fnref{fn2}}
\author[a]{M. Fransen}
\author[a,c]{A. Grelli}
\author[a,b]{I. Hobus}
\author[a,c]{A. Isakov}
\author[d]{A. Junique}
\author[d]{P. Leitao}
\author[d]{M. Mager}
\author[d]{Y. Otarid}
\author[d]{F. Piro\fnref{fn3}}
\author[a,c]{M.J. Rossewij}
\author[a,c]{M. Selina}
\author[a,b]{S. Solokhin}
\author[d]{W. Snoeys}
\author[d]{N. Tiltmann}
\author[a]{A. Vitkovskiy}
\author[a,b]{H. Wennlöf}
\author{on behalf of the ALICE collaboration}

%\email{jory.sonneveld@nikhef.nl}
%% Author affiliation
\affiliation[a]{organization={Nikhef},%Department and Organization
           addressline={Science Park 105}, 
           city={Amsterdam},
           postcode={1098XG}, 
           country={The Netherlands}}
\affiliation[b]{organization={University of Amsterdam},%Department and Organization
           addressline={Science Park 904}, 
           city={Amsterdam},
           postcode={1098XH}, 
           country={The Netherlands}}
\affiliation[c]{organization={Utrecht University},%Department and Organization
           addressline={Princetonplein 5}, 
           city={Utrecht},
           postcode={3584CC}, 
           country={The Netherlands}}
\affiliation[d]{organization={CERN},%Department and Organization
           addressline={CH-1211}, 
           postcode={Geneve 23}, 
           country={Switzerland}}

%\affiliation[a]{Nikhef, Science Park 105, 1098 XG Amsterdam, the Netherlands}
%\affiliation[b]{CERN, CH-1211 Geneve 23, Switzerland}
%\affiliation[c]{Utrecht University, Princetonplein 1, 3584 CC Utrecht, the Netherlands}
%T\affiliation[d]{University of Amsterdam, Amsterdam, Netherlands}
\fntext[fn1]{Now with University of Geneva, Switzerland.}
\fntext[fn2]{Now with Brookhaven National Laboratory, US.}
\fntext[fn3]{Now with Miromico IC, Zurich, Switzerland.}

%% Abstract
\begin{abstract}
%% Text of abstract

In the LHC long shutdown 3, the ALICE experiment upgrades the inner layers of its Inner Tracker System with three layers of wafer-scale stitched sensors bent around the beam pipe. Two stitched sensor evaluation structures, the MOnolithic Stitched Sensor (MOSS) and MOnolithic Stitched Sensor with Timing (MOST) allow the study of yield dependence on circuit density, power supply segmentation, stitching demonstration for power and data transmission, performance dependence on reverse bias, charge collection performance, parameter uniformity across the chip, and performance of wafer-scale data transmission.

The MOST measures 25.9 cm x 0.25 cm, has more than 900,000 pixels of 18x18 $\mu$m$^2$ and emphasizes the validation of pixel circuitry with maximum density, together with a high number of power domains separated by switches allowing to disconnect faulty circuits. It employs 1 Gb/s 26 cm long data transmission using asynchronous, data-driven readout. This readout preserves information on pixel address, time of arrival and time over threshold. In the MOSS, by contrast, regions with different in-pixel densities are implemented to study yield dependence and are read synchronously.

MOST test results validated the concept of power domain switching and the data transmission over 26 cm stitched lines which are to be employed on the full-size, full-functionality ITS3 prototype sensor, MOSAIX. Jitter of this transmission is still under study. This proceeding summarizes the performance of the stitched sensor test structures with emphasis on the MOST.

\end{abstract}

%%Graphical abstract
%\begin{graphicalabstract}
%\includegraphics{grabs}
%\end{graphicalabstract}

%%Research highlights
%\begin{highlights}
%\item Research highlight 1
%\item Research highlight 2
%\end{highlights}

%% Keywords
\begin{keyword}
%% keywords here, in the form: keyword \sep keyword
monolithic
\sep stitched sensor
\sep pixel detectors
\sep silicon
\sep ALICE
\sep timing
%% PACS codes here, in the form: \PACS code \sep code

%% MSC codes here, in the form: \MSC code \sep code
%% or \MSC[2008] code \sep code (2000 is the default)

\end{keyword}

\end{frontmatter}

%% Add \usepackage{lineno} before \begin{document} and uncomment 
%% following line to enable line numbers
%% \linenumbers

%% main text
%%

%% Use \section commands to start a section
\section{The ALICE ITS3 Upgrade}
\label{its3}
In 2028, the ALICE experiment \cite{alice} at the CERN LHC \cite{LHC} plans to upgrade its inner tracking system (ITS) with 3 layers of 266 mm long wafer-scale half-layer monolithic stitched pixel sensors thinned to 50 $\mu$m, which are bent around the beam pipe and held in place by carbon foam \cite{its3tdr}. The innermost layer will be 58.7 mm wide and will be located at only 19 mm from the interaction point. It will have a radiation length of X/X$_0 \approx 0.09$ \%, only a quarter of the already very low material pixel detector now in operation \cite{its2tdr, ls2upgrade} (see N. Valle, these proceedings). In collaboration with CERN Experimental Physics R\&D in Engineering Run 1 (ER1) two prototypes have been produced with the Tower Partners Semiconductor Co., Ltd. (TPSco) 65 nm Image Sensor CMOS (ISC) Technology that have been tested for high energy physics \cite{dpts, apts, processmod} to learn about stitching, study sensor yield, and for general R\&D. The final prototype sensor, the MOnolithic Stitched Active pIXel (MOSAIX) \cite{leitao_development_2024}, is to include features of these prototypes, the MOSS (see F. Carnesecchi, these proceedings) \cite{leitao_development_2024, terlizzi_characterization_2024} and the MOST \cite{mariiaproceedings}.

%% Use \subsection commands to start a subsection.
%\subsection{Example Subsection}
%\label{subsec1}

%Subsection text.

%% Use \subsubsection, \paragraph, \subparagraph commands to 
%% start 3rd, 4th and 5th level sections.
%% Refer following link for more details.
%% https://en.wikibooks.org/wiki/LaTeX/Document_Structure#Sectioning_commands
\section{Monolithic Stitched Sensor with Timing: MOST}
\label{most}
\begin{figure*}[htbp]%% placement specifier
%% Use \includegraphics command to insert graphic files. Place graphics files in 
%% working directory.
\centering%% For centre alignment of image.
\includegraphics[width=0.8\textwidth, trim={10 50 50 40}, clip]{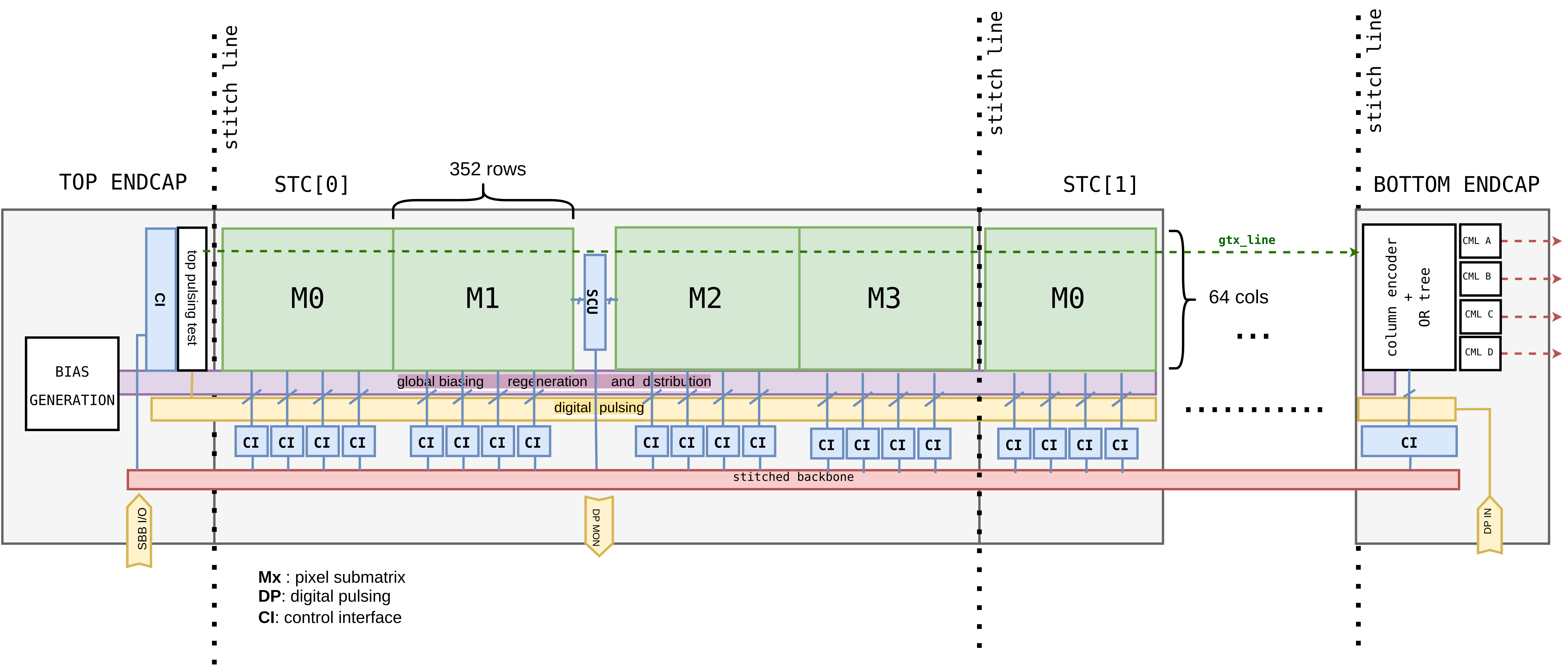}
\includegraphics[width=0.7\textwidth]{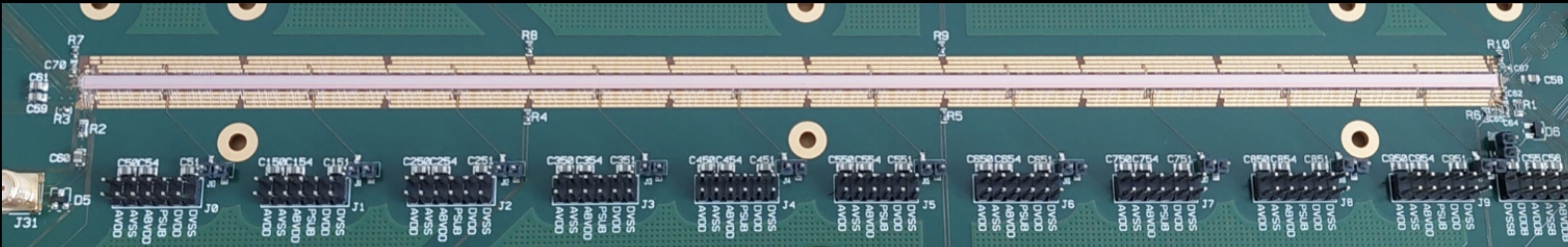}
%% Use \caption command for figure caption and label.
\caption{Top: top level diagram of the Monolithic Stitched Sensor with Timing. Bottom: Picture of a MOST bonded to a carrier board.}\label{fig:mostschematic}
%% https://en.wikibooks.org/wiki/LaTeX/Importing_Graphics#Importing_external_graphics
\end{figure*}

The MOST has 10 repeated stitched units (RSUs) with 90112 pixels each, and 352 rows and 64 columns of 18 $\mu$m pitch pixels in each RSU submatrix M0, M1, M2 and M3 (see Fig \ref{fig:mostschematic}). The main differences with the MOSS are a more granular powering scheme, sensor biasing via the front-end, and an asynchronous readout including Time over Threshold information. The MOST has global analog and digital power networks with switches dividing the chip in small fractions that can individually be connected or disconnected in case of shorts or defects. In the MOSS, only 20 power domains are accessible from separate pads. The MOST was made for testing of higher integration density in each pixel, detaching subsets of pixels with power gating to handle possible shorts, and the transmission speed of $\sim$ 1 Gbit/s over 26 cm of silicon. Currently, about 50 sensors are bonded, where the handling (picking, placement and gluing) of these thin chips of 259 $\times$ 2.5 mm$^2$ is extremely challenging.

\section{Granular Power Gating}
\label{powermost}
\begin{figure}[htbp]%% placement specifier
%% Use \includegraphics command to insert graphic files. Place graphics files in 
%% working directory.
\centering%% For centre alignment of image.
\includegraphics[width=\columnwidth]{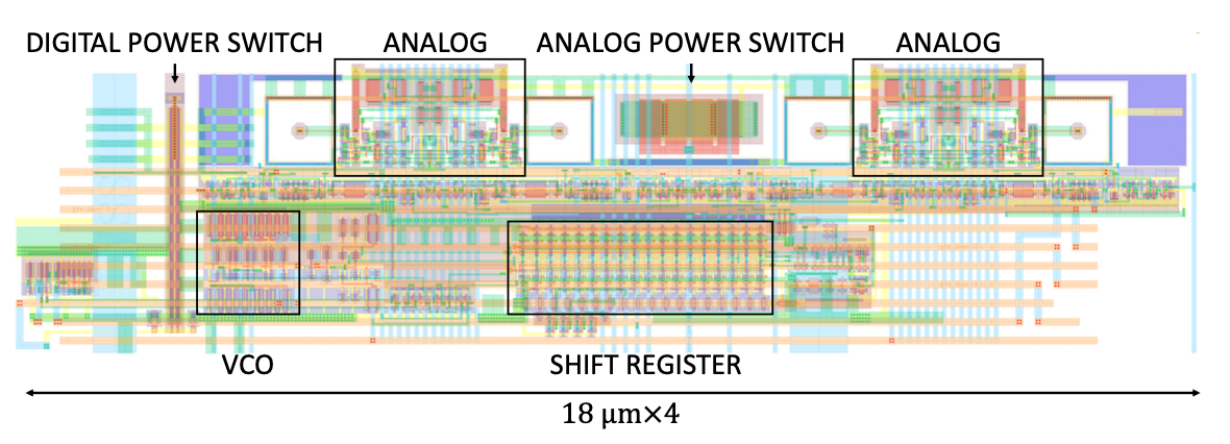}
%% Use \caption command for figure caption and label.
\caption{Four MOST pixels. A maximally dense pixel design has been adopted behind the digital and analog power switches. Figure from \cite{mariiaproceedings}.}\label{fig:pixeldesign}
%% https://en.wikibooks.org/wiki/LaTeX/Importing_Graphics#Importing_external_graphics
\end{figure}

The MOST design contains analog power gating controlling 4 rows of 64 pixels per RSU (2560 switches) and digital power gating for 352 pixels in a column per RSU (3520 switches). %, allowing for a maximally dense design after the power switches (see Fig. \ref{fig:pixeldesign}). 
 Shorts or defects can then be isolated and be prevented from affecting the full chip, allowing the internal circuitry powered through the power switches to be designed at the maximum density defined by the design rules (see Fig. \ref{fig:pixeldesign}).
However, this requires maintaining performance when switching off a certain region, and the observed high pixel yield demonstrates that we can afford to design the pixel circuitry controlled by the power switches at full density. The variation of the threshold was found to remain well below the noise level of 15 e$^-$ for pixels where a neighboring region was switched off, showing that the power segmentation to be used in the ITS3 is fully functional. With an expected MIP signal of 600 e$^-$ in the MOST, the lowest threshold found for this sensor was around 200 e$^-$, which can be further reduced with optimization of the chip operational settings.

\section{Yield of the MOST}
\label{yield}
\begin{figure}[htbp]%% placement specifier
\centering%% For centre alignment of image.
\includegraphics[width=0.92\columnwidth, trim={7 10 0 0}, clip]{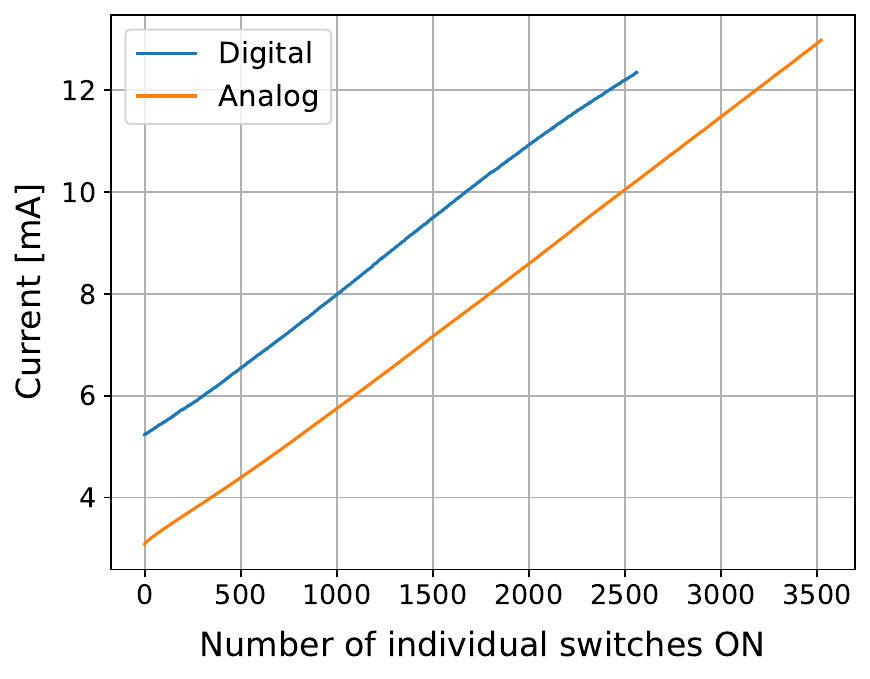}
\includegraphics[width=\columnwidth, trim={0 0 50 50}, clip]{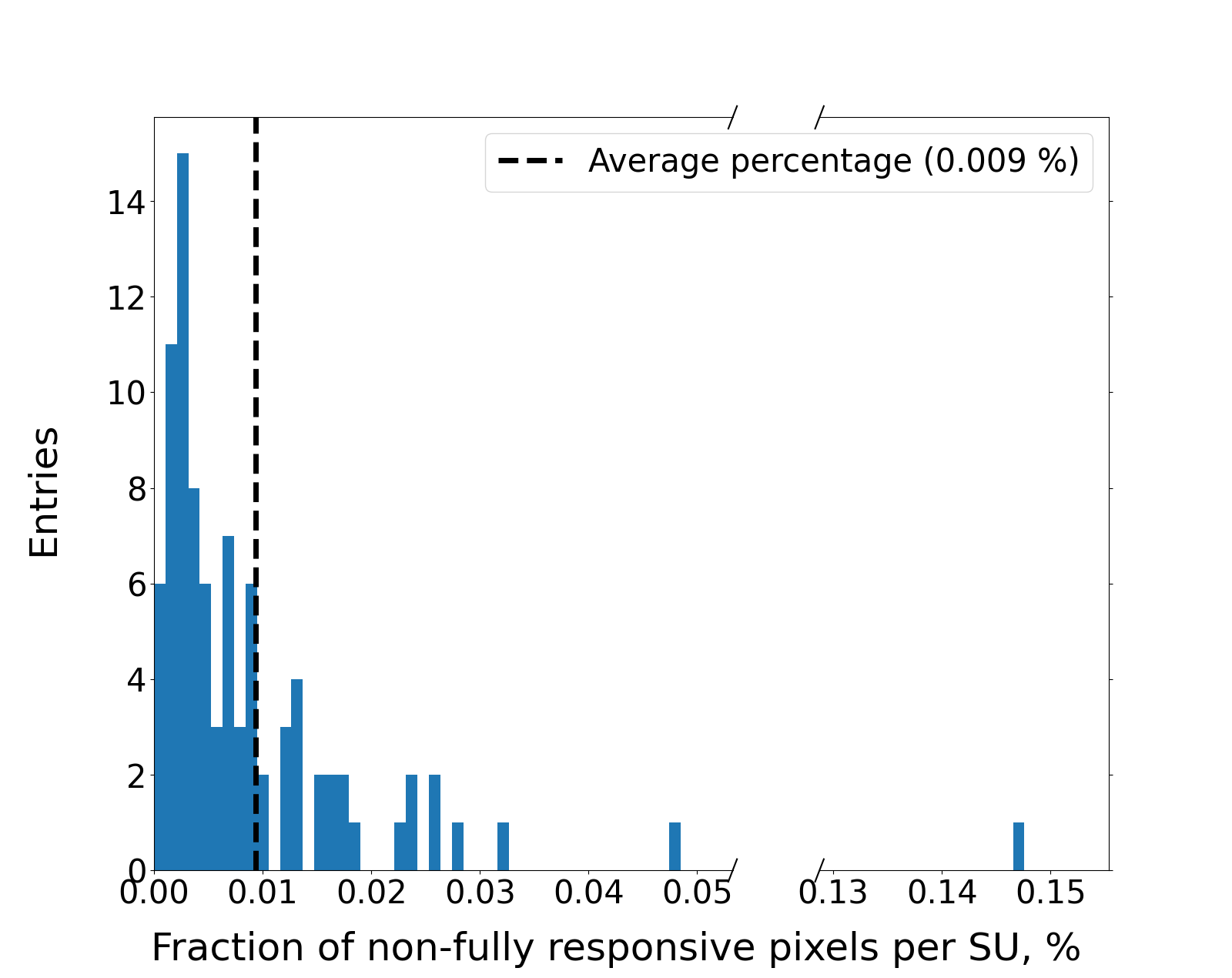}
\caption{The MOST power consumption scales as expected when turning on individual power gates (top, from \cite{mariiaproceedings}) and has a very good pixel response yield with an average of only 0.009\% of pixels per stitched unit (SU) not responding (bottom). }\label{fig:pixelyield}
\end{figure}

For a fraction of the MOST sensors that cannot be powered, shorts have been observed compatible with process issues already reported in the MOSS yield studies \cite{eberwein_yield_2024}. This is expected to be solved for the next run. For chips that can be powered, a nice linear increase in power is seen when switching on the digital and analog power gates sequentially. For these sensors, almost all pixels in the MOST respond to pulsing: a yield of more than 99.99\% was found on 80 tested RSUs (see Fig. \ref{fig:pixelyield}), with one outlier from 180 pixels in one stitched unit likely to come from a defect in one of the shared lines in the missing rows. This demonstrates the feasibility of dense pixel design for future applications.

\section{Reverse sensor bias}
\label{bias}
\begin{figure}[htbp]%% placement specifier
\centering%% For centre alignment of image.
\includegraphics[width=0.9\columnwidth]{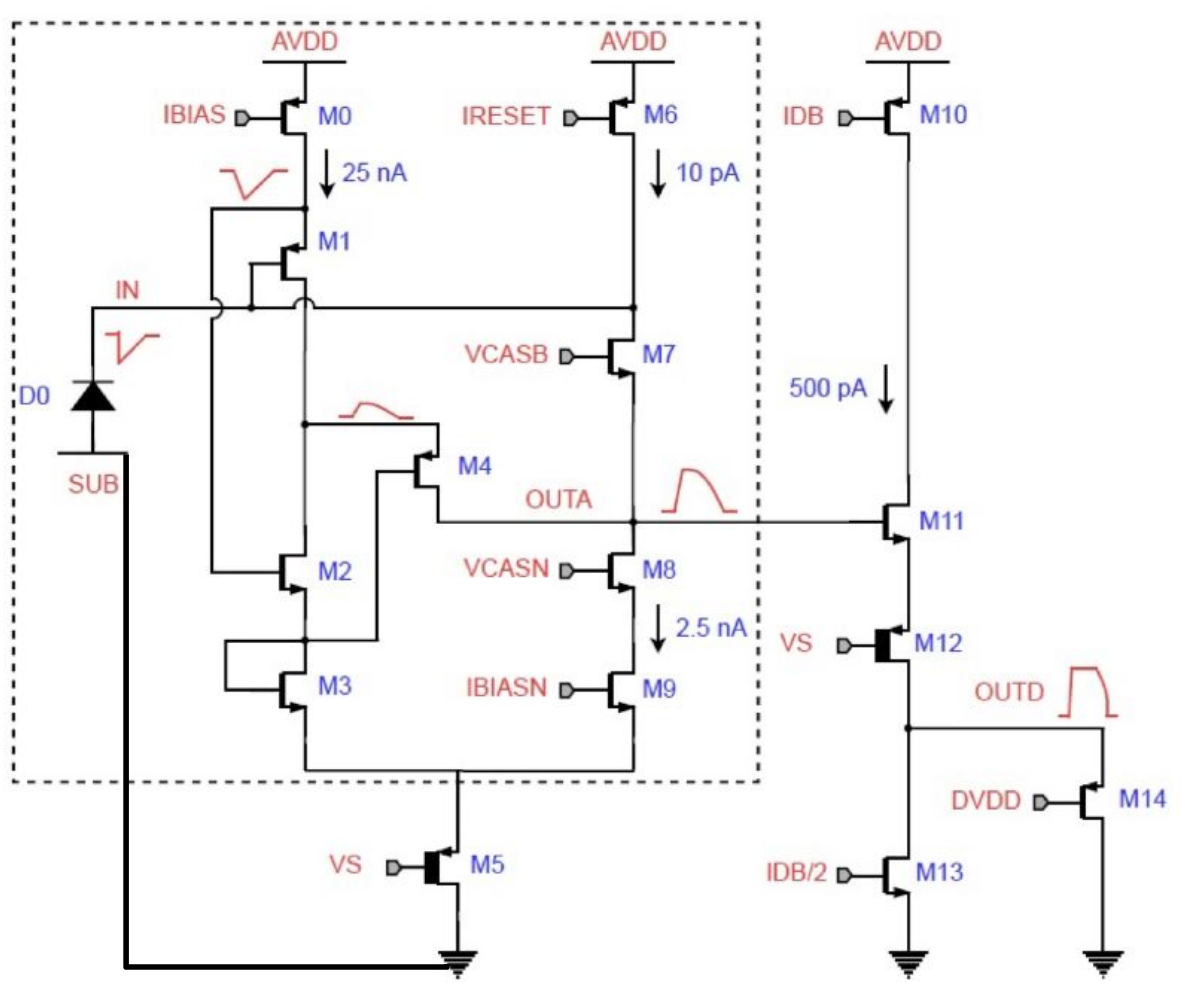}
\caption{Schematic of the MOST pixel front end. The analog ground is connected to the sensor substrate. The shifting voltage VS is used to control the reverse bias, where the analog power AVDD needs to be high enough to keep a margin. OUTD is the output hit signal of the pixel.}\label{fig:frontend}
\end{figure}

\begin{figure}[t]%% placement specifier
\centering%% For centre alignment of image.
\includegraphics[width=0.95\columnwidth, trim={0 16 0 0}, clip]{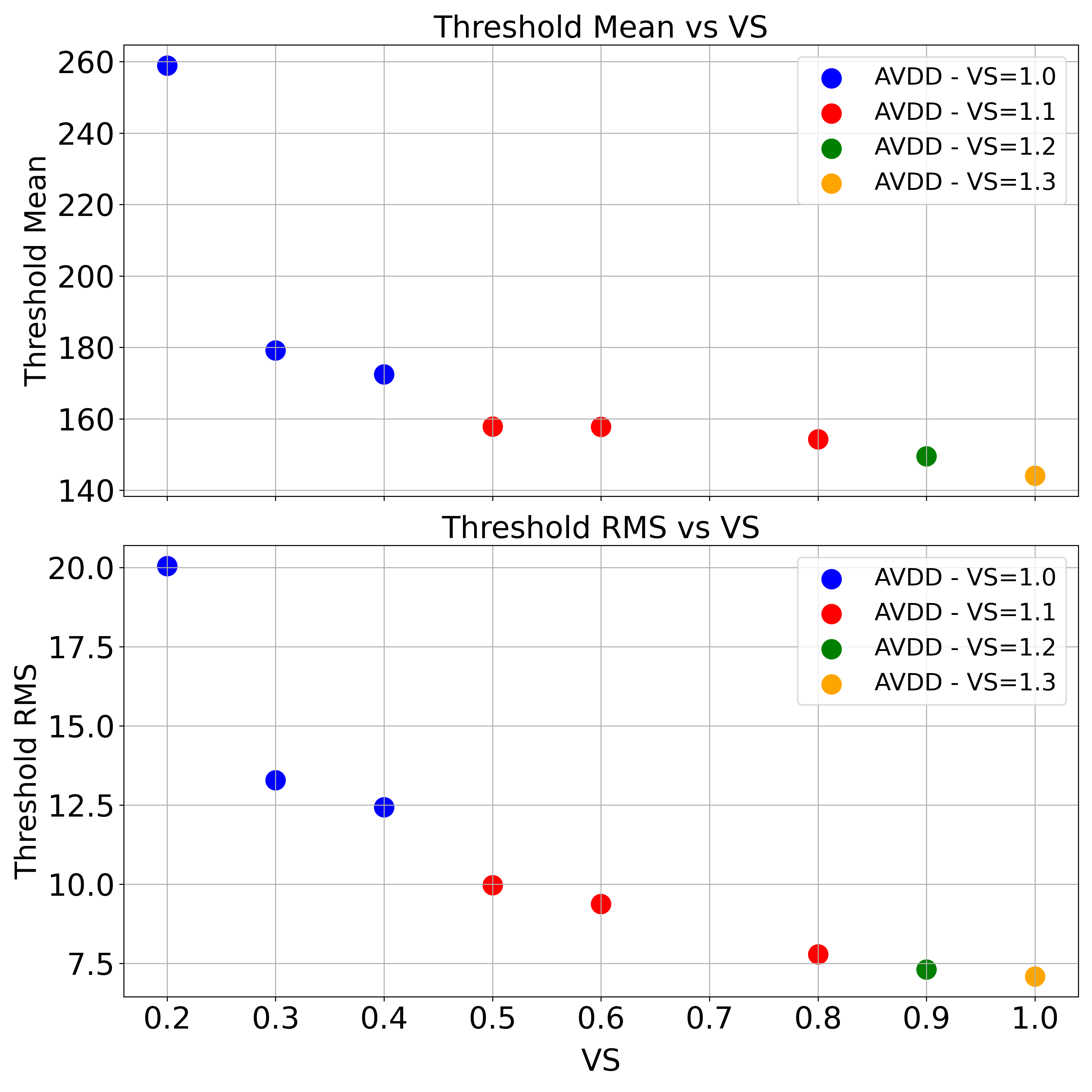}
\caption{Threshold and threshold spread as a function of sensor bias through a changing shifting voltage. The analog power is at a fixed difference with the shifting voltage, with colors indicating the different offsets. Figure from \cite{mariiaproceedings}.}\label{fig:sensorbias}
\end{figure}

The MOST has the analog circuit ground connected to the substrate, where a reverse sensor bias is then applied by shifting up the potential at the input of the front-end by varying a shifting voltage (VS). This allows for a simpler design where there is no need to account for reverse bias protection structures, but for a higher sensor bias the analog supply needs to be increased as well to provide enough headroom for the front-end circuitry, leading to a higher analog power consumption (see Fig. \ref{fig:frontend}). As expected, an increased VS results in a higher reverse bias, which in turn results in a lower sensor capacitance and a lower charge threshold and threshold spread (see Fig. \ref{fig:sensorbias}) \cite{mariiaproceedings}. It also results in lower noise, but other components are dominant and are now further optimized to reduce the noise below 15 e$^-$.

\section{Asynchronous readout and timing}
\label{timing}
\begin{figure}[h]%% placement specifier
\centering%% For centre alignment of image.
\includegraphics[width=\columnwidth]{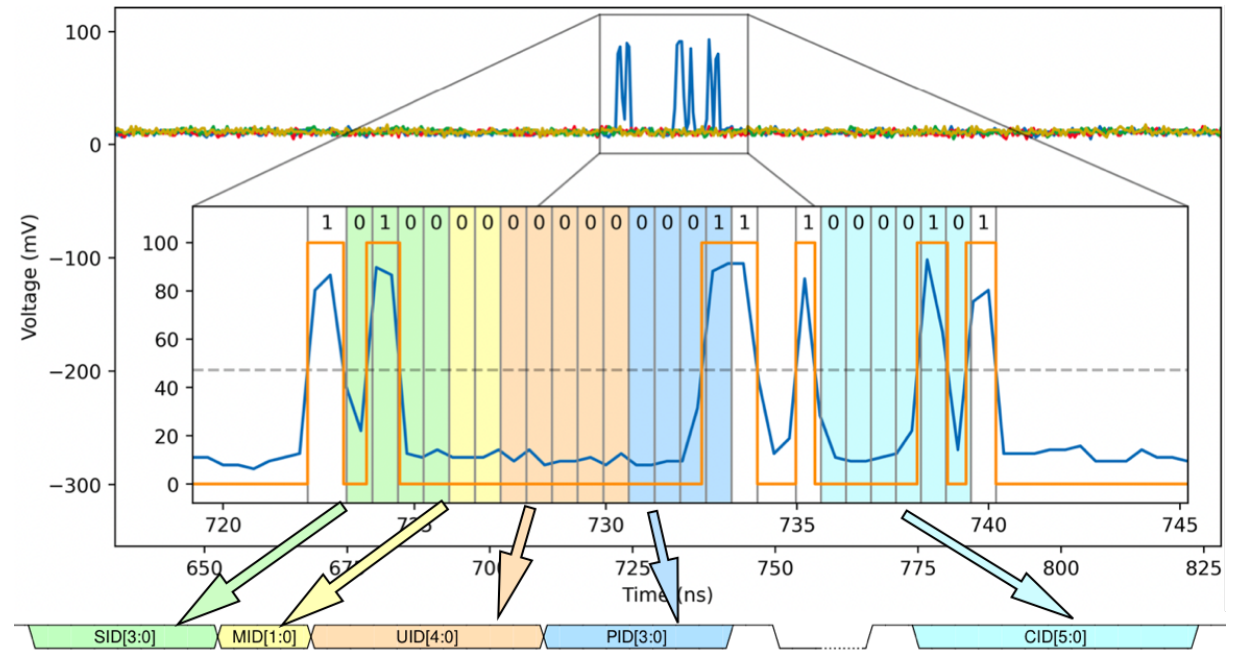}
\caption{Hit information from the MOST is encoded in a waveform containing the stitch ID, matrix ID, unit ID, pixel ID and column ID. Hit information is sent twice, once on the rising and once on the falling edge of a hit signal, to encode time over threshold information.}\label{fig:mosthit}
\end{figure}

The MOST has an asynchronous readout where direct hit information is sent out of the chip (see Fig. \ref{fig:mosthit}), and time over threshold information is obtained through sending hit information twice, when the comparator passes the threshold and when it returns to zero. This is similar to the readout scheme of a previously studied digital pixel test structure \cite{dpts}. Data is transmitted over the full sensor length, where the column address is added to the bit sequence in the chip periphery or end cap (see Fig. \ref{fig:mostschematic}). All hits go through an OR logic and are then transmitted through one of 4 current-mode logic (CML) outputs. To minimize collisions from neighboring pixel hits, each column has 4 transmission lines. 

\begin{figure}[htbp]%% placement specifier
\centering%% For centre alignment of image.
\includegraphics[width=0.9\columnwidth,  trim={0 0 0 87}, clip]{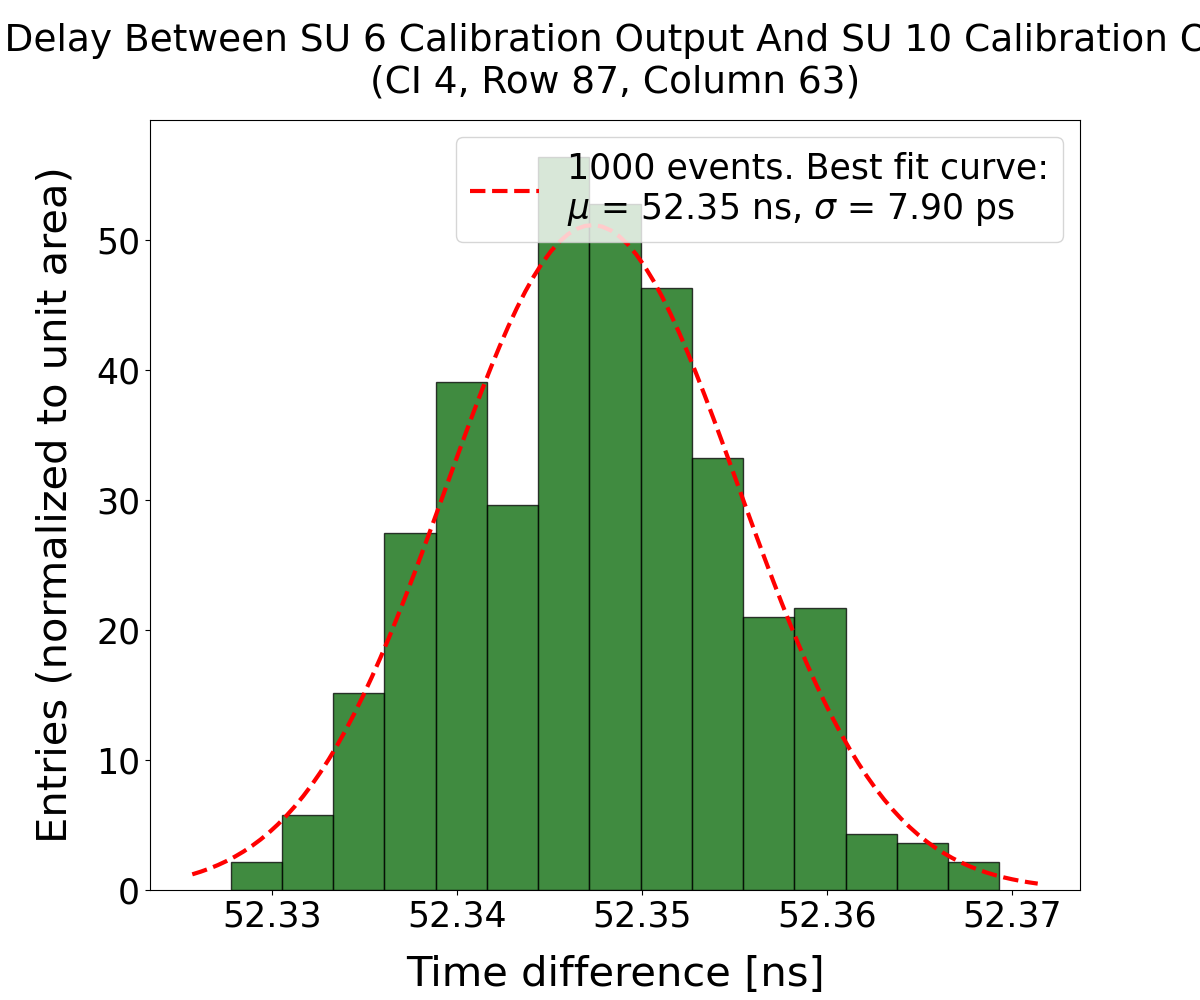}
\includegraphics[width=\columnwidth]{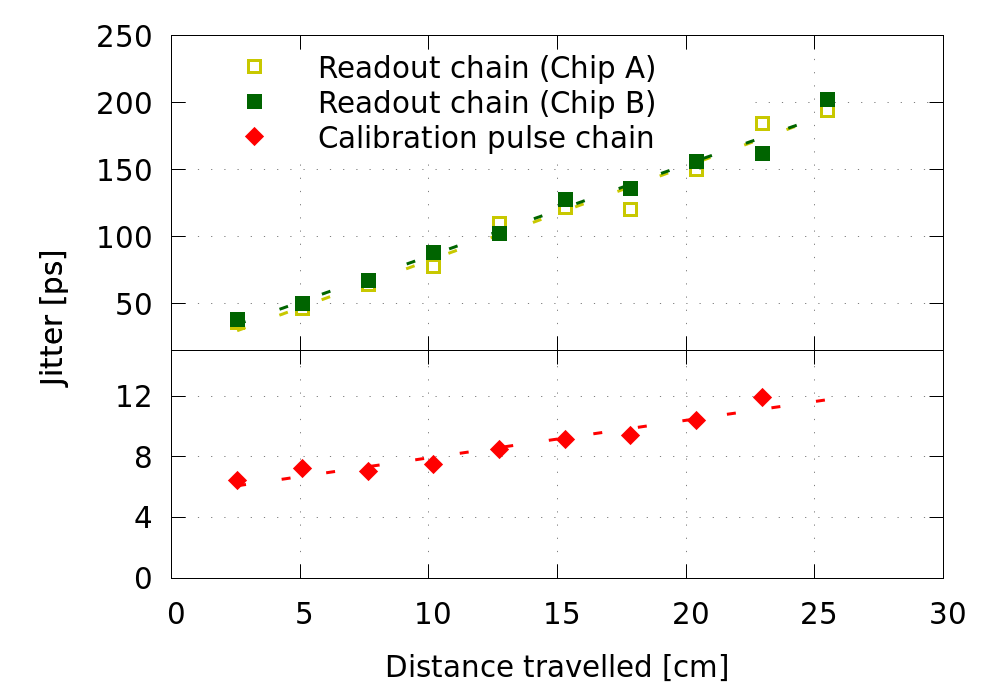}
\caption{Preliminary measurement of the time delay between the output of a calibration pulse on RSU 6 at about 10 cm along the sensor and the last RSU 10 for Control Interface 4, row 87 and column 63 (top). The spread indicates the jitter, which is about 8 ps. The jitter on the readout line is much larger than on the calibration pulse line, and increases linearly over the sensor distance (bottom). Figures from \cite{mariiaproceedings}.}\label{fig:jitter}
\end{figure}

If hit and timing information is preserved over the full chip length, a time to digital converter would not be needed for every pixel or group of pixels but can instead in future designs be integrated in the periphery of a chip. This would require low jitter in the leading edge of the signal. Preliminary measurements show a jitter of 8 ps on the calibration path of a pulse (see Fig. \ref{fig:jitter}). The separate readout path shows an order of magnitude higher jitter and a linear increase over the length of the sensor not compatible with a square-root increasing jitter from noise. Investigation into these differences are ongoing, but 8 ps jitter on the calibration pulse is encouraging for transmission of timing information over a 26 cm long stitched sensor.

\section{MOSAIX}
\label{mosaix}
\begin{figure}[htbp]%% placement specifier
\centering%% For centre alignment of image.
\includegraphics[width=\columnwidth, trim={256 0 70 0}, clip]{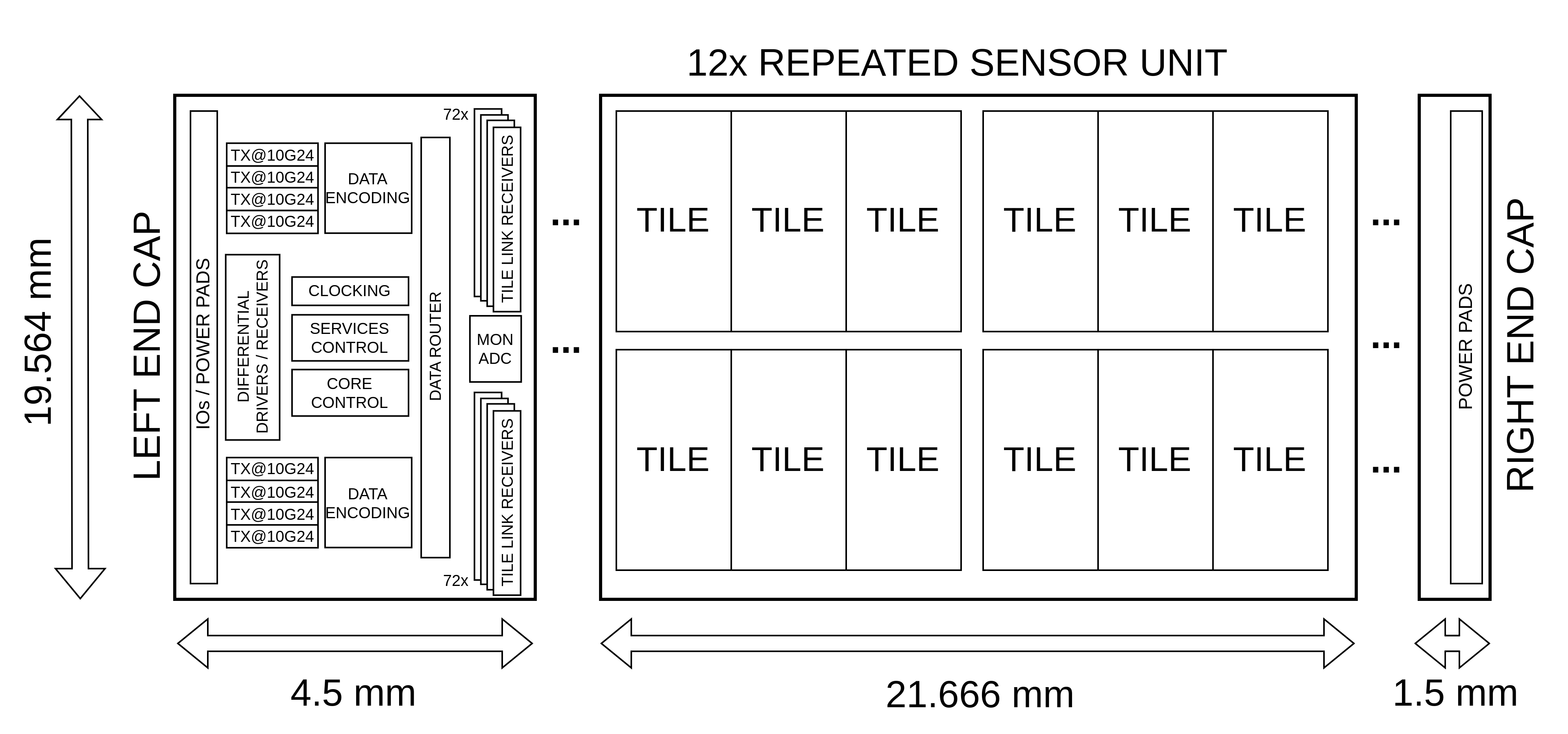}
\caption{Block diagram of the full-size full functionality prototype for the ITS3, the MOSAIX.}\label{fig:mosaix}
\end{figure}

The MOnolithic Stitched Active pIXel, or MOSAIX, is the full-size full functionality prototype for the ITS3, with a readout through the left end cap and powering through both end caps. It features 12 RSUs, with 12 "tiles" per RSU that each have independent powering, control, and readout (see Fig. \ref{fig:mosaix}). This results in 144 tiles, where each tile can be powered individually and covers 0.7\% of the sensor area, which is a larger fraction than for the MOST but a much smaller fraction than for the MOSS. The MOSAIX will then also have a less dense design than the MOST. Excluding the endcaps, the sensitive area of the MOSAIX will be 93\%.

\section{Conclusion and Outlook}
\label{conclusion}
The promising results of the MOST give a path to a more dense circuit design and long-distance on-chip transmission with timing information. Its powering scheme that is to be used in ITS3 works as designed, and the MOST has a pixel response yield of more than 99.98 \%. The front end  is fully functional and its reverse bias scheme is working as expected. Though the timing performance is currently still under investigation, the asynchronous readout of the MOST is fully functional.

%% Use a table environment to create tables.
%% Refer following link for more details.
%% https://en.wikibooks.org/wiki/LaTeX/Tables
%\begin{table}[t]%% placement specifier
%% Use tabular environment to tag the tabular data.
%% https://en.wikibooks.org/wiki/LaTeX/Tables#The_tabular_environment
%\centering%% For centre alignment of tabular.
%\begin{tabular}{l c r}%% Table column specifiers
%% Tabular cells are separated by &
%  1 & 2 & 3 \\ %% A tabular row ends with \\
%  4 & 5 & 6 \\
%  7 & 8 & 9 \\
%\end{tabular}
%% Use \caption command for table caption and label.
%\caption{Table Caption}\label{fig1}
%\end{table}

%% The Appendices part is started with the command \appendix;
%% appendix sections are then done as normal sections
%\appendix
%\section{Example Appendix Section}
%\label{app1}

%Appendix text.

%% For citations use: 
%%       \cite{<label>} ==> [1]

%%
%Example citation, See \cite{lamport94}.

%% If you have bib database file and want bibtex to generate the
%% bibitems, please use
%%
%%  \bibliographystyle{elsarticle-num} 
%%  \bibliography{<your bibdatabase>}

%% else use the following coding to input the bibitems directly in the
%% TeX file.

%% Refer following link for more details about bibliography and citations.
%% https://en.wikibooks.org/wiki/LaTeX/Bibliography_Management

\end{document}